\documentclass[draft]{agujournal2019}
\usepackage{url} 
\usepackage{lineno}
\usepackage[inline]{trackchanges} 
\usepackage{soul}

\usepackage{amsmath,amssymb}
\usepackage{graphicx}
\usepackage{booktabs}
\usepackage{float}
\usepackage{tabularx}
\usepackage{adjustbox}
\usepackage{siunitx}
\usepackage{cuted}
\usepackage{placeins}
\usepackage{caption}
\usepackage{subcaption}
\usepackage{xcolor}

\draftfalse

\begin{document}

%
%


\title{An Adaptive Spatiotemporal Clustering Framework for 3D Ocean Subsurface Temperature Reconstruction}

%
%




\authors{Ming Shan Loo\affil{1}, Wengen Li\affil{1}, Xudong Jiang\affil{1}, Hailiang Cheng\affil{2}, Zhifei Zhang\affil{3}, Jihong Guan\affil{1}, and Yichao Zhang\affil{1}}


\affiliation{1}{School of Computer Science and Technology, Tongji University, Shanghai, China}
\affiliation{2}{The Key Laboratory of Road and Traffic Engineering, Ministry of Education, College of Transportation, Tongji University, Shanghai, China}
\affiliation{3}{Project Management Office of China National Scientific Seafloor Observatory, Tongji University, Shanghai, China}




\correspondingauthor{Wengen Li}{lwengen@tongji.edu.cn}



\begin{keypoints}
\item An adaptive spatiotemporal clustering framework is proposed for reconstructing 3D ocean thermal structures using remote sensing data.

\item The framework captures vertical thermocline connectivity and temporal variability of ocean subsurface temperature through clustering.

\item The framework can seamlessly integrate with various deep learning models to improve 3D temperature reconstruction from surface data.

\end{keypoints}

%
%

%
%


\begin{abstract}
The reconstruction of ocean subsurface temperature (OST) using satellite remote sensing data holds significant scientific value for advancing the understanding of ocean dynamics and climate variability. However, the scarcity of subsurface observations, combined with the high degree of nonlinearity and spatiotemporal heterogeneity in subsurface processes, poses substantial challenges to the accuracy and generalization capability of traditional reconstruction methods. To address these limitations, this study proposes an adaptive framework that could capture both vertical structural dependencies and temporal variation patterns of OST via spatio-temporal clustering. By incorporating this framework with various deep learning models, e.g., dual-path convolutional neural networks (DP-CNN), Attention U-Net, and Vision Transformer (ViT), the OST field can be accurately reconstructed at a global scale only using surface observations, i.e., sea surface temperature (SST), sea surface salinity (SSS), sea surface height (SSH), and sea surface wind (SSW). Experimental results demonstrate that multiple deep learning methods using the proposed framework largely outperform their original counterparts, yielding improvements in RMSE ranging from 12.4\% to 27.2\%. This study provides a reliable solution for subsurface temperature reconstruction, offering important implications for meteorological modeling and climate change assessment.
\end{abstract}

\section{Introduction}

Ocean subsurface temperature (OST) serves as a key indicator of oceanic heat content and vertical structure, playing a crucial role in processes such as thermocline variation, ocean circulation, and ocean-atmosphere energy exchange~\cite{stewart2008introduction,bindoff2022changing}. However, direct in-situ observations of subsurface temperature remain limited in spatial and temporal coverage because of the high cost and sparse distribution of observational platforms such as Argo floats and moored buoys~\cite{roemmich2009argo,abraham2013review,von2013monitoring,zhou2023high}. This observational gap highlights the importance of subsurface temperature reconstruction, which offers a viable pathway to obtaining spatially and temporally complete subsurface thermal fields for advancing climate prediction and ecosystem monitoring~\cite{cutolo2024cloinet}.

Meanwhile, satellite remote sensing provides abundant high-resolution surface information~\cite{ali2004estimation,yang2023spatial}, including sea surface temperature (SST), salinity (SSS), height (SSH) and wind (SSW). Although these variables do not directly reflect subsurface conditions, they carry latent correlations with vertical oceanic structures~\cite{song2022inversion}.
Leveraging such surface-subsurface correlations, a growing body of research has focused on reconstructing 3D subsurface temperature fields from satellite-derived surface data.

Current reconstruction methods generally fall into numerical models and data-driven models, the latter including traditional machine learning and deep learning. Numerical models (e.g., ROMS~\cite{shchepetkin2005regional}, HYCOM~\cite{chassignet2007hycom}) reconstruct subsurface thermal fields by simulating ocean dynamics based on governing equations, achieving high physical consistency but at the cost of intensive computation and high sensitivity to initial and boundary conditions. Traditional machine learning approaches, such as support vector machine (SVM)~\cite{su2015estimation}, random forests (RF)~\cite{su2018retrieving,su2021predicting}, and XGBoost~\cite{su2019estimating}, learn empirical relationships from data with high computational efficiency, yet their limited model capacity constrains the representation of complex non-linear patterns in ocean data. Deep learning models~\cite{su2021predicting,xie2022reconstruction,mao2023reconstructing,wu2024pgtransnet,zheng2024generating,jiang2025spatio,11214178,xie2025dual,yu2025reconstruction} have demonstrated promising performance in modeling non-linear mappings between surface and subsurface variables, but often treat the ocean as a homogeneous system, insufficiently accounting for the spatial and temporal heterogeneity of subsurface thermal fields.

\begin{figure*}[t]
\centering
\begin{subfigure}[t]{0.58\textwidth}
    \centering
    \includegraphics[width=\linewidth]{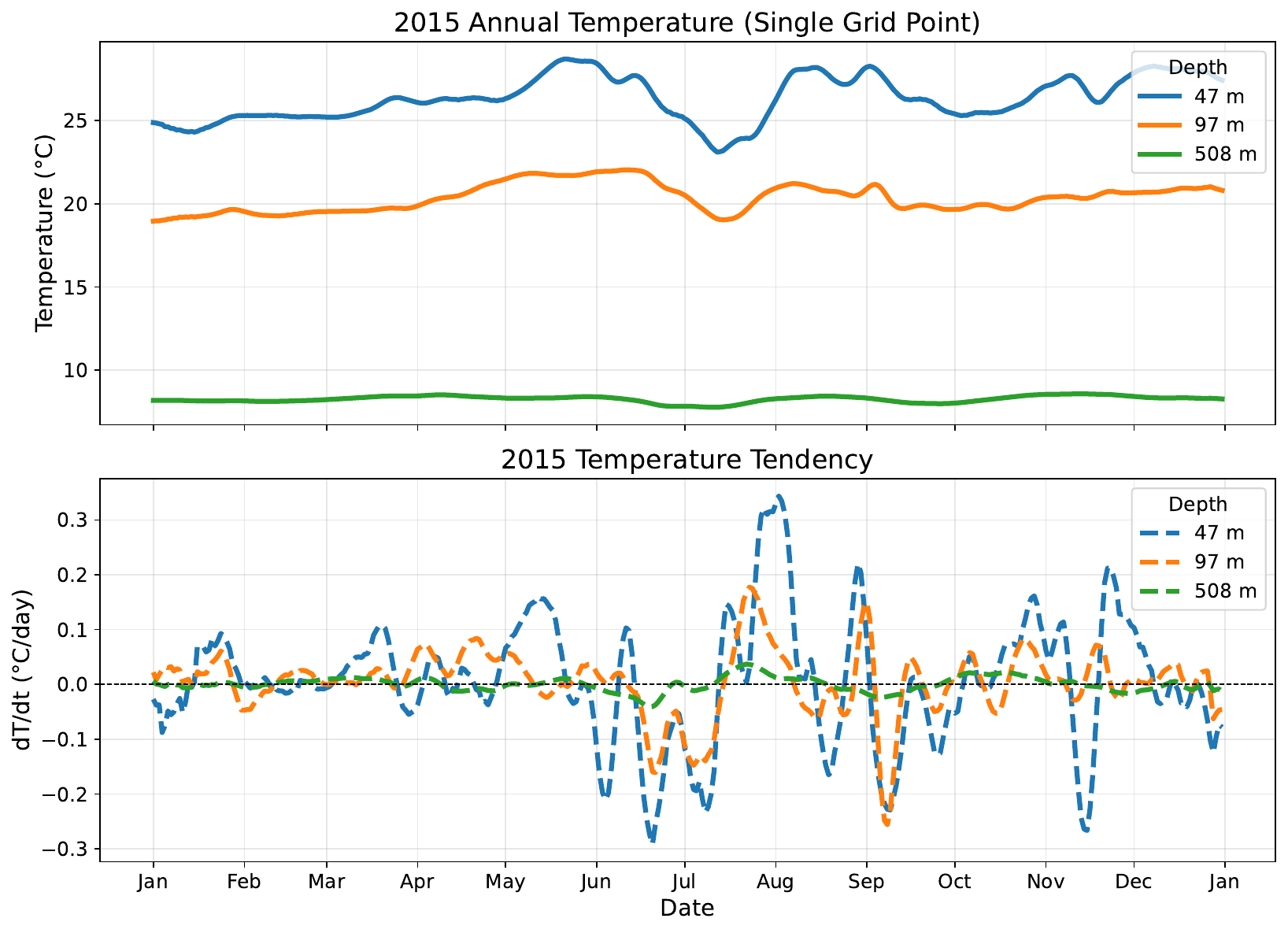}
    \caption{Annual temperature variations at selected depths.}
    \label{fig:annual_temp}
\end{subfigure}
\hfill
\begin{subfigure}[t]{0.36\textwidth}
    \centering
    \includegraphics[width=\linewidth]{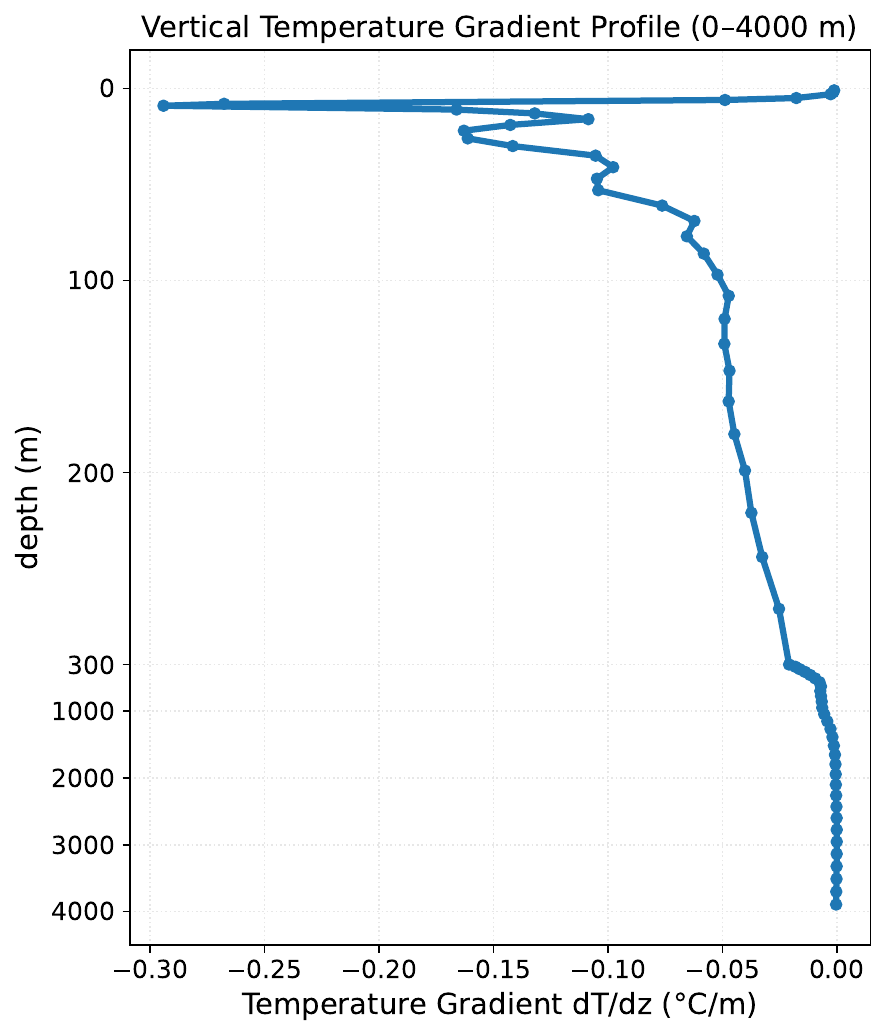}
    \caption{Vertical temperature gradient profile.}
    \label{fig:vertical_gradient}
\end{subfigure}
\caption{Comparison of temporal and vertical characteristics of ocean temperature at a single point (12.50$^\circ$, 112.50$^\circ$). (a) Annual cycle at multiple depths. (b) Vertical gradient structure.}
\label{fig:annual_vs_gradient}
\end{figure*}

This heterogeneity is clearly illustrated in Figure~\ref{fig:annual_vs_gradient}. In the temporal dimension, the upper-layer temperature demonstrates a clear annual cycle driven by seasonal forcing, with considerable differences in phase and amplitude among different depths. In the vertical spatial dimension, the temperature gradient reveals a distinct stratification: a strong thermocline and rapid attenuation are observed near the surface, whereas deeper waters remain relatively stable. These observations suggest that effective reconstruction requires jointly modeling vertical dependencies and temporal dynamics, rather than treating the ocean as a uniform system.

To address these challenges, we propose an adaptive spatio-temporal clustering framework for three-dimensional subsurface temperature reconstruction. By analyzing both vertical and temporal patterns in the data, our framework segments the ocean into coherent spatio-temporal subregions, enabling the training of localized deep learning models tailored to the unique structural characteristics of each subregion. This design significantly improves reconstruction accuracy and model generalizability. Moreover, the proposed framework is architecture-agnostic and can be seamlessly integrated with typical neural network backbones, including DP-CNN, Attention-UNet, ViT, FFNN, LSTM, and OCNN.

Our experimental results across multiple model configurations demonstrate the effectiveness and robustness of this clustering-guided strategy. This work not only contributes a practical tool for subsurface temperature reconstruction but also provides methodological insights into the handling of heterogeneity in spatiotemporal oceanographic modeling.

In sum, the contributions of this work include: 
\begin{itemize}
\item We design an adaptive framework that integrates spatiotemporal clustering with data-driven reconstruction. Unlike conventional approaches that assume global homogeneity, our method partitions the field into coherent vertical and temporal sub-blocks, reducing heterogeneity and enabling more precise subsurface temperature reconstruction. The framework is modular and can flexibly incorporate diverse neural architectures as reconstruction backbones.
\item We conduct experiments on two representative tropical ocean regions using nearly three decades of reanalysis data (1993-2022). The results show that the framework achieves consistent improvements across six representative neural architectures, confirming its generality and effectiveness.
\end{itemize}

The rest of this work is organized as below. Section~\ref{sec:relatedwork} reviews the related work.
Section~\ref{sec:Study Areas and Data} introduces the study areas and datasets used in this work. 
Section~\ref{sec:Methodology} describes the proposed methodology, including a framework overview, vertical dependency clustering, temporal dynamics clustering, and the final 3D reconstruction strategy. 
Section~\ref{sec:Experiments and Results} presents the experiments and results, covering the experimental settings, comparison with baseline models, vertical and temporal clustering results, and an ablation study. 
Section~\ref{sec:Visualization and Discussion} provides visualization results and further discussion. 
Finally, Section~\ref{sec:Conclusion} concludes the paper and outlines future research directions.

\section{Related Work}\label{sec:relatedwork}
Existing studies on subsurface temperature reconstruction can be roughly divided into two categories, i.e., numerical models and data-driven methods.

\textbf{Numerical Models.} Numerical ocean models, such as the Regional Ocean Modeling System (ROMS)~\cite{shchepetkin2005regional} and the HYbrid Coordinate Ocean Model (HYCOM)~\cite{chassignet2007hycom}, provide physically interpretable estimates of subsurface temperature fields by solving governing equations of ocean dynamics. These models can capture large-scale circulation patterns and thermohaline structures. However, they are highly sensitive to initial and boundary conditions~\cite{santana2022data,rahaman2023impact}, require a large number of tunable parameters, and often involve prohibitive computational costs and a long simulation time, which limit their scalability for real-time or large-domain applications. 
To mitigate the limitations of purely physics-based models, hybrid approaches combine physical simulations with data-driven corrections or constraints. For instance, data assimilation techniques integrate in-situ and satellite observations into numerical models, improving reconstruction accuracy by leveraging both physical consistency and observational data~\cite{carton2008reanalysis,cutolo2024cloinet}. Nonetheless, such methods still inherit the high complexity, computational burden, and parameter sensitivity of physical models.

\textbf{Data-driven methods.}  
Data-driven methods include traditional machine learning methods and deep learning methods. Traditional machine learning methods, including SVM~\cite{su2015estimation}, Random Forests (RF)~\cite{su2018retrieving,su2021predicting}, and XGBoost~\cite{su2019estimating}, have been employed to learn statistical mappings between ocean surface observations and subsurface temperature fields. While these methods are computationally efficient and easy to implement, they struggle to capture the nonlinear, high-dimensional, and spatiotemporally dependent nature of ocean processes, and typically rely on extensive manual feature engineering, which limits their generalization capability across diverse oceanic regimes. 
Recent advances in deep learning have further expanded the potential of data-driven modeling. Architectures such as Feedforward Neural Network(FFNN)~\cite{tian2022reconstructing}, Convolutional Neural Networks (CNNs)~\cite{mao2023reconstructing,smith2023reconstruction}, Long Short-Term Memory (LSTM) networks~\cite{buongiorno2020deep,su2021predicting}, Attention U-Nets~\cite{xie2022reconstruction}, and Vision Transformers (ViTs)~\cite{wu2024pgtransnet,xie2025dual,yu2025reconstruction} have been successfully introduced into subsurface temperature reconstruction tasks. These models excel at capturing complex spatial patterns and temporal dependencies from high-dimensional inputs without the need for manual feature engineering. However, most existing deep learning models treat the ocean as a spatially homogeneous medium, neglecting the inherent spatial heterogeneity and temporal heterogeneity of subsurface thermal structures which are critical for accurate 3D reconstruction. 

\textbf{Ocean Regionalization and Clustering-based Modeling.}
In recent years, clustering algorithms have been increasingly introduced to address the spatiotemporal heterogeneity of ocean processes by exploiting the intrinsic similarity within oceanographic data. In the field of ocean pattern recognition, clustering methods have proven effective in identifying coherent oceanic regimes. For instance, Sun et al. applied the K-means algorithm to delineate hydrographic regions in the Antarctic Ocean~\cite{sun2021clustering}, while Radin et al. employed a similar approach to extract climate regions characterized by shared variability patterns~\cite{radin2024unveiling}.
Beyond pattern recognition, clustering techniques have also been adopted in ocean field reconstruction, where the core idea remains consistent: leveraging the intrinsic similarity within oceanographic data to reduce complexity and improve modeling performance. The general strategy is to partition the ocean into subregions with similar dynamical characteristics and then train independent models within each subregion. Lu et al. employed K-means clustering to divide the global ocean into several static thermal provinces, thereby providing more targeted local training data for neural networks~\cite{lu2019subsurface}. Fan et al. integrated K-means clustering with a Transformer architecture, extracting multi-scale dynamical features to reduce reconstruction errors of subsurface temperatures in the western Pacific~\cite{fan2024reconstruction}. Luo et al. combined clustering with ensemble learning methods, grouping multi-source sea surface variables and spatiotemporal coordinates using Gaussian Mixture Models (GMM) or K-means to yield additional improvements in reconstruction performance~\cite{luo2025subsurface}.

Despite the effectiveness of these methods, existing clustering-based approaches still present certain limitations in capturing the 3D spatiotemporal variability of ocean processes. From a spatial perspective, most studies focus primarily on two-dimensional horizontal partitioning and do not explicitly model vertical dependencies or thermal structures across different depth layers. From a temporal perspective, many existing methods rely on static clustering results and neglect the temporal heterogeneity associated with seasonal and interannual variability. Even in studies where temporal information (e.g., months) is incorporated into the feature space~\cite{luo2025subsurface}, it is merely used as one of several input variables for multivariate clustering, rather than being specifically designed to capture the temporal evolution of ocean temperature within particular cycles. These limitations suggest that the vertical structure of subsurface thermal fields and their temporal evolution have not yet been fully explored.

\section{Study Areas and Data}\label{sec:Study Areas and Data}

\begin{figure}[!h]
\centering
\includegraphics[width=0.8\columnwidth]{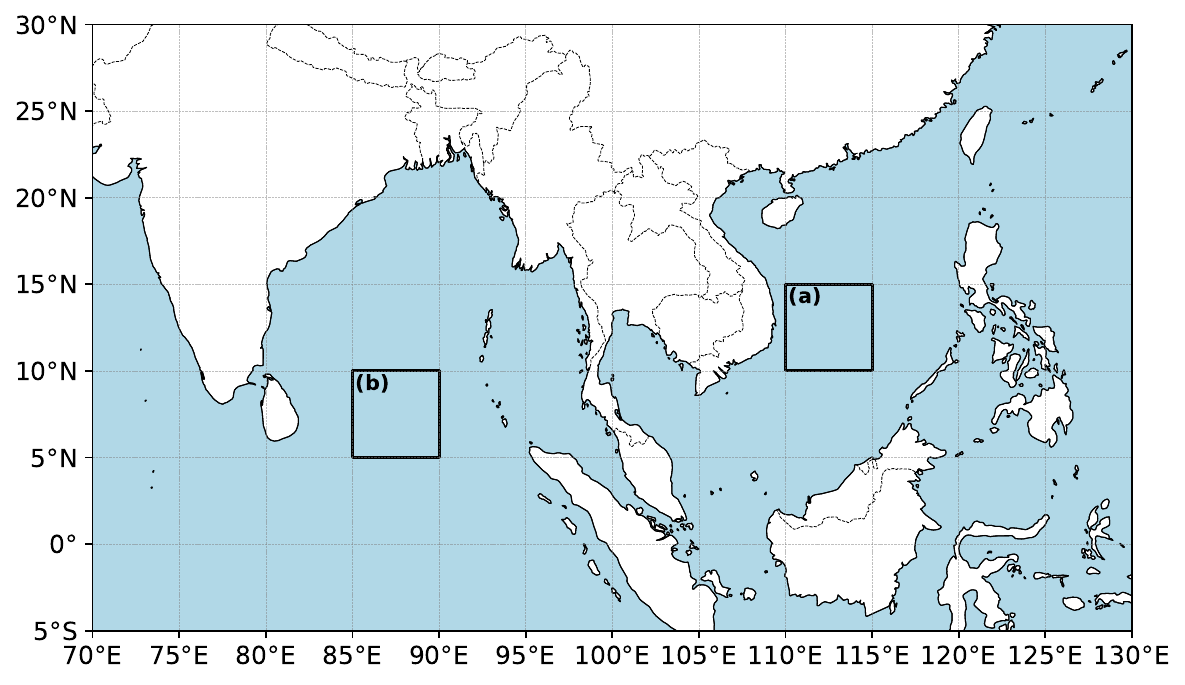}
\caption{Two study areas in the Indian Ocean and South China Sea, respectively.}
\label{fig:study_regions}
\end{figure}

This study focuses on two representative tropical ocean areas: (a) the northwestern sector of the South China Sea (110°E-115°E, 10°N-15°N), and (b) a central equatorial sector of the Indian Ocean (85°E-90°E, 5°N-10°N).  

The datasets used in this work include both sea surface observations and subsurface oceanic reanalysis products. All variables were regridded to a uniform spatial resolution of 0.25$^{\circ} \times$0.25$^{\circ}$ to ensure consistency across sources. The temporal coverage spans 29 years (1993--2022), providing long-term continuity and reliability for model training. 

The surface input variables employed in this study include sea surface temperature (SST), sea surface salinity (SSS), sea surface height (SSH), and sea surface wind (SSW).  
The SST~\cite{CMEMS_SST,sst_worsfold2024presenting}, SSS~\cite{CMEMS_SSS,sss1_droghei2016combining,sss2_nardelli2016multi,sss3_droghei2018new,sss4_sammartino2022retrieving}, and SSH~\cite{CMEMS_GREP} data were obtained from the E.U. Copernicus Marine Service Information (CMEMS) reanalysis products.  
Their original spatial resolutions are 0.05$^{\circ} \times$0.05$^{\circ}$ (SST), 0.125$^{\circ} \times$0.125$^{\circ}$ (SSS), and 0.25$^{\circ} \times$0.25$^{\circ}$ (SSH), respectively.  
The SSW data~\cite{ncar_gdex_dataset_d745001} were obtained from the Cross-Calibrated Multi-Platform Ocean Surface Wind Vector Analysis Product V2 provided by the NSF NCAR Geoscience Data Exchange, with an original spatial resolution of 0.25$^{\circ} \times$0.25$^{\circ}$.

\begin{figure}[h]
\centering
\includegraphics[width=\linewidth]{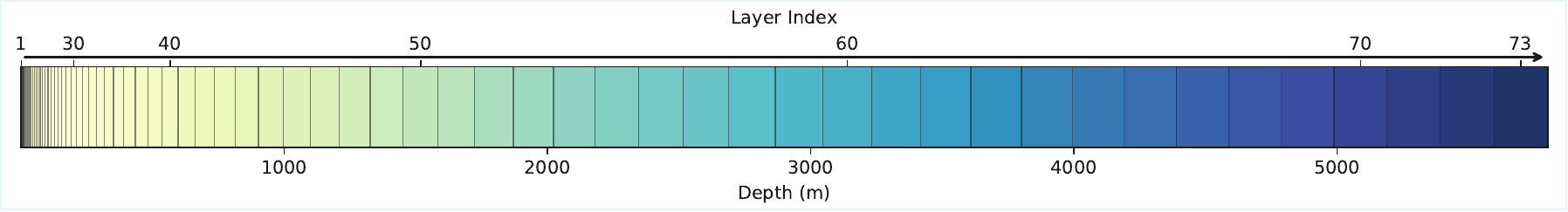}
\caption{
    Non-uniform vertical stratification of the subsurface temperature datasets.
    The vertical coordinate is discretized into 73 layers with irregular spacing.
    The detailed depth boundaries (in meters) are:
    1, 2, 3, 5, 6, 8, 9, 11, 13, 16, 19, 22, 26, 30, 35, 41, 47, 53, 61, 69, 77, 86, 97, 108, 120, 133, 147, 163, 180, 199, 221, 244, 271, 300, 333, 370, 411, 457, 508, 565, 628, 697, 773, 856, 947, 1045, 1151, 1265, 1387, 1516, 1652, 1795, 1945, 2101, 2262, 2429, 2600, 2776, 2955, 3138, 3324, 3513, 3704, 3897, 4093, 4289, 4488, 4687, 4888, 5089, 5291, 5494, and 5698.
}
\label{fig:stratification}
\end{figure}

The OST data~\cite{CMEMS_GREP} were obtained from the CMEMS reanalysis product with an original spatial resolution of 0.25$^{\circ} \times$0.25$^{\circ}$. As illustrated in Figure~\ref{fig:stratification}, the datasets include 73 vertical layers, spanning from the sea surface to a maximum depth of 5,698 m, providing a detailed depiction of the ocean's vertical thermal structure.

\section{Methodology}\label{sec:Methodology}

\subsection{Framework Overview}
\begin{figure*}[htbp]
    \centering
    \includegraphics[width=\textwidth]{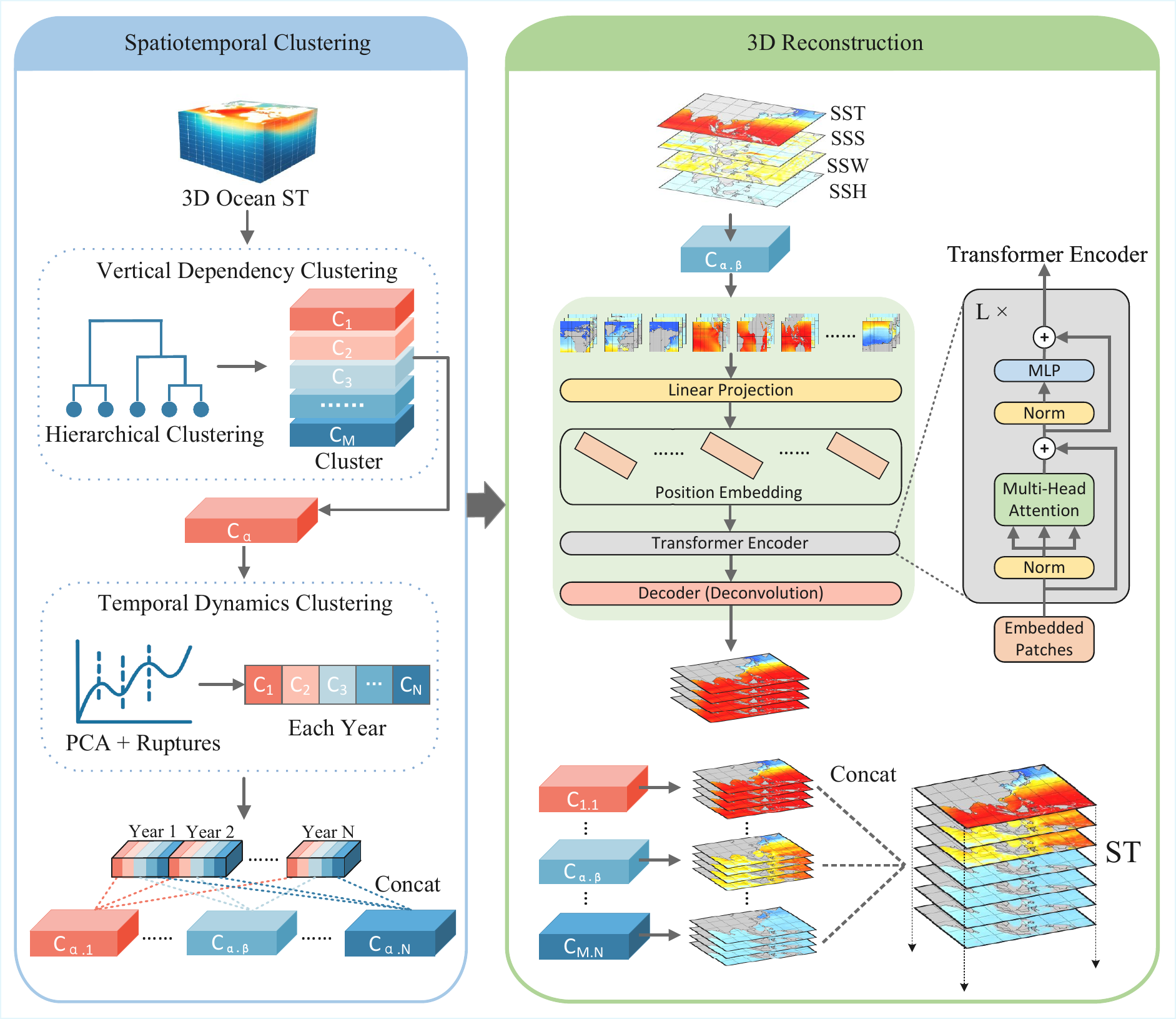}
    \caption{
        Overview of the proposed framework which consists of two stages, i.e., spatiotemporal clustering and 3D reconstruction.
        First, the clustering module partitions the 3D ocean data into sub-blocks based on depth structures and temporal characteristics, effectively capturing vertical dependencies and temporal dynamics.
        Then, the 3D reconstruction module independently feeds each clustered sub-block into the backbone models to reconstruct subsurface temperatures.
        The framework supports various neural network architectures, thereby offering strong flexibility and adaptability.
        Finally, the reconstructed outputs are merged to form a complete subsurface temperature field.
    }
    \label{fig:framework}
\end{figure*}

Figure~\ref{fig:framework} illustrates the proposed adaptive OST reconstruction framework that consists of two stages, i.e., spatiotemporal clustering and 3D reconstruction. 

First, the spatiotemporal clustering involves the structural segmentation of the three-dimensional ocean temperature field to enhance the model’s ability to capture spatiotemporal heterogeneity. This stage consists of two key components: a vertical dependency clustering module, which extracts structural correlations across depth layers, and a temporal dynamics clustering module, which identifies non-stationary patterns over time.

The second stage is 3D reconstruction that performs independent modeling and prediction for each spatiotemporal sub-block derived from the pre-clustering stage. In this stage, each spatiotemporal sub-block obtained from the pre-clustering process is independently modeled and reconstructed. The input to the model consists of multi-source remote sensing observations, including SST, SSS, SSW, and SSH. A Vision Transformer (ViT) model is employed to illustrate the idea of learning the mapping between surface variables and subsurface thermal structures for each sub-block. Finally, the reconstructed outputs of all sub-blocks are aggregated to recover the complete three-dimensional subsurface temperature field, thereby enabling accurate estimation of the ocean’s thermal structure.

\subsection{Vertical Dependency Clustering}

To enhance the model’s sensitivity to vertical structure, we design a vertical dependency clustering module that captures layer-wise structural coherence in the ocean temperature field. This module aims to identify physically coherent depth intervals where temperature variations exhibit strong correlation, thereby improving the model's capacity to learn meaningful vertical representations.

We begin by computing a temporally averaged three-dimensional temperature field:
\begin{equation}
\bar{T}(x, y, z) = \frac{1}{T} \sum_{t=1}^{T} T(x, y, z, t)
\end{equation}
where $T(x, y, z, t)$ denotes temperature at spatial location $(x, y, z)$ and time $t$, and $\bar{T}$ captures the static vertical structure.

To derive a single vertical profile for clustering, we apply a spatial average across the horizontal domain:
\begin{equation}
\bar{T}(z) = \frac{1}{N_x N_y} \sum_{x=1}^{N_x} \sum_{y=1}^{N_y} \bar{T}(x, y, z)
\end{equation}
This results in a depth-wise temperature vector $\bar{T}(z) \in \mathbb{R}^{D}$, where $D$ is the total number of depth levels.

Next, hierarchical agglomerative clustering is employed on the vertical profile $\bar{T}(z)$, using pairwise distances (e.g., Euclidean) to construct a dendrogram:
\begin{equation}
\text{dist}(i, j) = \| \bar{T}(z_i) - \bar{T}(z_j) \|
\end{equation}
The tree is cut at a predefined number of clusters $K$,
yielding $K$ vertical clusters:
\begin{equation}
\mathcal{Z} = \{ \mathcal{Z}_1, \mathcal{Z}_2, \dots, \mathcal{Z}_K \}
\end{equation}
Each cluster $\mathcal{Z}_k$ consists of depth layers that are contiguous and thermodynamically consistent.

By aligning model training with these vertical segments, we enable a structurally modular learning strategy that better captures depth-dependent variability, leading to improved generalization and reconstruction accuracy in subsurface temperature modeling.

\subsection{Temporal Dynamics Clustering}

To enhance the model's ability to capture non-stationary temporal dynamics in the ocean temperature field, we introduce a temporal dynamics clustering module that segments the time domain based on dominant seasonal transitions. To mitigate interannual noise and highlight consistent intra-annual patterns, we compute a \textit{climatological average} over multiple years to construct a typical year temperature sequence, i.e., 
\begin{equation}
T_{\text{typical}}(d) = \frac{1}{N} \sum_{y=1}^{N} T_y(d), \quad d = 1, 2, ..., D
\end{equation}
where $T_y(d)$ denotes the temperature on day $d$ of year $y$, and $N$ is the total number of years. This results in a representative annual cycle that captures stable seasonal evolution.

We then apply principal component analysis (PCA) to each vertically clustered sub-block to extract dominant temporal variation modes from the typical-year sequence. Let $\mathbf{X} \in \mathbb{R}^{D \times H}$ represent the temperature series of $H$ horizontal grid cells over $D$ days, the first principal component $\mathbf{p}_1$ is computed by:
\begin{equation}
\mathbf{p}_1 = \arg\max_{\|\mathbf{w}\|=1} \| \mathbf{Xw} \|^2
\end{equation}
where $\mathbf{w} \in \mathbb{R}^H$ is a unit vector representing the projection direction across the $H$ horizontal grid cells.


Subsequently, we employ change-point detection via dynamic programming 
on $\mathbf{p}_1$ to locate abrupt shifts in temporal trends. These breakpoints delineate sub-periods of homogeneous temporal dynamics.

The resulting segments define \textit{temporal clusters}, which are combined with vertical clusters to form spatiotemporal sub-blocks. These blocks serve as independent training units for downstream temperature field reconstruction.
This strategy enhances the model’s ability to learn temporally localized features, reduces interference from mixed climatic signals, and improves both generalization and reconstruction accuracy.

\subsection{3D Reconstruction}

The 3D reconstruction module is designed to perform independent modeling and temperature field reconstruction for each spatiotemporal sub-block. Its core objective is to reconstruct the 3D subsurface temperature structure of the ocean using multi-source surface remote sensing observations, including SST, SSS, SSW, and SSH.

Formally, the surface variables are defined as:
\begin{equation}
\mathbf{X}_{s}(t, \mathbf{r}) = \{\mathrm{SST}(t, \mathbf{r}), \mathrm{SSS}(t, \mathbf{r}), \mathrm{SSW}(t, \mathbf{r}), \mathrm{SSH}(t, \mathbf{r})\}
\end{equation}
where $t$ denotes time and $\mathbf{r}$ represents the horizontal spatial coordinates. The reconstruction model aims to learn the mapping:
\begin{equation}
f_{\theta}: \mathbf{X}_{s}(t, \mathbf{r}) \longrightarrow T(z, t, \mathbf{r})
\end{equation}
with $T(z, t, \mathbf{r})$ representing the subsurface temperature distribution at depth $z$, and $\theta$ denoting the trainable parameters of the backbone network.

Specifically, each sub-block generated from the spatiotemporal clustering stage is independently input into a selected backbone model for training and prediction. This strategy avoids the interference caused by spatiotemporal heterogeneity and allows the model to focus on learning localized patterns.

For each sub-block $B_{\alpha,\beta}$ determined by vertical cluster $\alpha \in \{1,2,\dots,M\}$ and temporal cluster $\beta \in \{1,2,\dots,N\}$, the reconstruction is expressed as:
\begin{equation}
\widehat{T}^{(\alpha,\beta)}(z,t,\mathbf{r}) = f_{\theta_{\alpha,\beta}}(\mathbf{X}_s(t,\mathbf{r})), \quad (z,t,\mathbf{r}) \in B_{\alpha,\beta}
\end{equation}
During training, supervised learning is employed to capture the complex nonlinear mapping between surface conditions and vertical thermal structure. The optimization objective is defined as:
\begin{equation}
\mathcal{L}(\theta_{\alpha,\beta}) = \frac{1}{N_{\alpha,\beta}} \sum_{i=1}^{N_{\alpha,\beta}} \| f_{\theta_{\alpha,\beta}}(\mathbf{X}_{s,i}) - T_{i}(z,t,\mathbf{r}) \|_{2}^{2}
\end{equation}
where $N_{\alpha,\beta}$ is the number of training samples in sub-block $B_{\alpha,\beta}$.

After reconstruction, all sub-block outputs are spatially merged according to their original positions to recover the complete 3D subsurface temperature field:
\begin{equation}
\widehat{T}(z,t,\mathbf{r}) = \mathcal{M}\Big(\{\widehat{T}^{(\alpha,\beta)}(z,t,\mathbf{r}) \;|\; (z,t,\mathbf{r}) \in B_{\alpha,\beta}, \; \alpha=1,\dots,M, \; \beta=1,\dots,N\}\Big)
\end{equation}
where $\mathcal{M}(\cdot)$ denotes the merging operator that integrates predictions from all sub-blocks.

This block-wise modeling strategy improves the model’s representation ability in heterogeneous regions and reduces the impact of cross-region information interference, thereby enhancing the overall reconstruction accuracy and generalization performance.
  
To further illustrate the 3D reconstruction process, we take ViT as an example backbone model. Unlike convolutional networks that extract features through local receptive fields, ViT first partitions each spatiotemporal sub-block $\mathbf{X}_{s}(t,\mathbf{r})$ into a set of non-overlapping image patches. Each patch is flattened and projected into a $d$-dimensional embedding space, forming a sequence of tokens:
\begin{equation}
\mathbf{Z}_0 = [\mathbf{x}_1\mathbf{E}; \mathbf{x}_2\mathbf{E}; \dots; \mathbf{x}_P\mathbf{E}] + \mathbf{E}_{pos}
\end{equation}
where $\mathbf{x}_i$ denotes the $i$-th patch, $\mathbf{E}$ is a learnable embedding matrix, $P$ is the total number of patches, and $\mathbf{E}_{pos}$ represents positional embeddings to preserve spatial order information.  

These tokens are then processed by multiple Transformer encoder layers, where the self-attention mechanism captures long-range dependencies across spatial positions:
\begin{equation}
\text{Attention}(\mathbf{Q}, \mathbf{K}, \mathbf{V}) = \text{Softmax}\left(\frac{\mathbf{QK}^\top}{\sqrt{d}}\right)\mathbf{V}
\end{equation}
with $\mathbf{Q}$, $\mathbf{K}$, and $\mathbf{V}$ denoting the query, key, and value matrices projected from the input tokens $\mathbf{Z}_0$. The multi-head attention mechanism allows the model to simultaneously aggregate contextual information from different spatial regions.  

Finally, the Transformer output representation $\mathbf{H}$ is fed into a lightweight decoder $g_{\phi}$ implemented with deconvolution (transposed convolution) layers, which progressively upsample the latent features into the predicted subsurface temperature field $\widehat{T}^{(\alpha,\beta)}(z,t,\mathbf{r})$:
\noindent
\begin{equation}
\begin{aligned}
\widehat{T}^{(\alpha,\beta)}(z,t,\mathbf{r})
= g_{\phi}(\mathbf{H}),
\quad (z,t,\mathbf{r}) \in B_{\alpha,\beta}, \;
\alpha = 1,\dots,M, \;
\beta = 1,\dots,N
\end{aligned}
\end{equation}
where $\phi$ denotes the learnable parameters of the decoder.

\section{Experiments and Results}\label{sec:Experiments and Results}
\subsection{Experiment Settings}

In our experiments, each dataset was partitioned into training, validation, and testing subsets with a ratio of 8:1:1, where the split was performed globally to ensure independence across subsets and to avoid data leakage. The training set was used to fit the model parameters, the validation set for hyperparameter tuning and early stopping, and the testing set was reserved for the final evaluation of model performance.  
Root mean square error (RMSE) and mean absolute error (MAE) were adopted as the evaluation metrics, i.e.,   
\begin{equation}
\mathrm{RMSE} = \sqrt{\frac{1}{N}\sum_{i=1}^{N}(y_i - \hat{y}_i)^{2}}
\label{eq:rmse}
\end{equation}

\begin{equation}
\mathrm{MAE} = \frac{1}{N}\sum_{i=1}^{N}\lvert y_i - \hat{y}_i \rvert
\label{eq:mae}
\end{equation}
where $y_i$ and $\hat{y}_i$ denote the observed and predicted values, respectively, and $N$ is the total number of samples. MAE provides a more direct measure of the average prediction deviation, while RMSE penalizes larger errors more heavily. The combination of both metrics offers a more comprehensive evaluation of reconstruction performance.

During training, the number of epochs was set to $300$, with a batch size of $32$. The Adam optimizer was employed with an initial learning rate of $1\times10^{-2}$, which was dynamically adjusted using a decay factor of $0.01$. An early stopping strategy with a patience of 10 epochs was applied to prevent overfitting. All experiments were conducted on a workstation equipped with an NVIDIA RTX 3090 GPU, and identical configurations were maintained across runs to ensure fairness and reproducibility.

\begin{figure*}[htbp]
    \centering
    \begin{subfigure}[t]{0.48\textwidth}
        \centering
        \includegraphics[width=\textwidth]{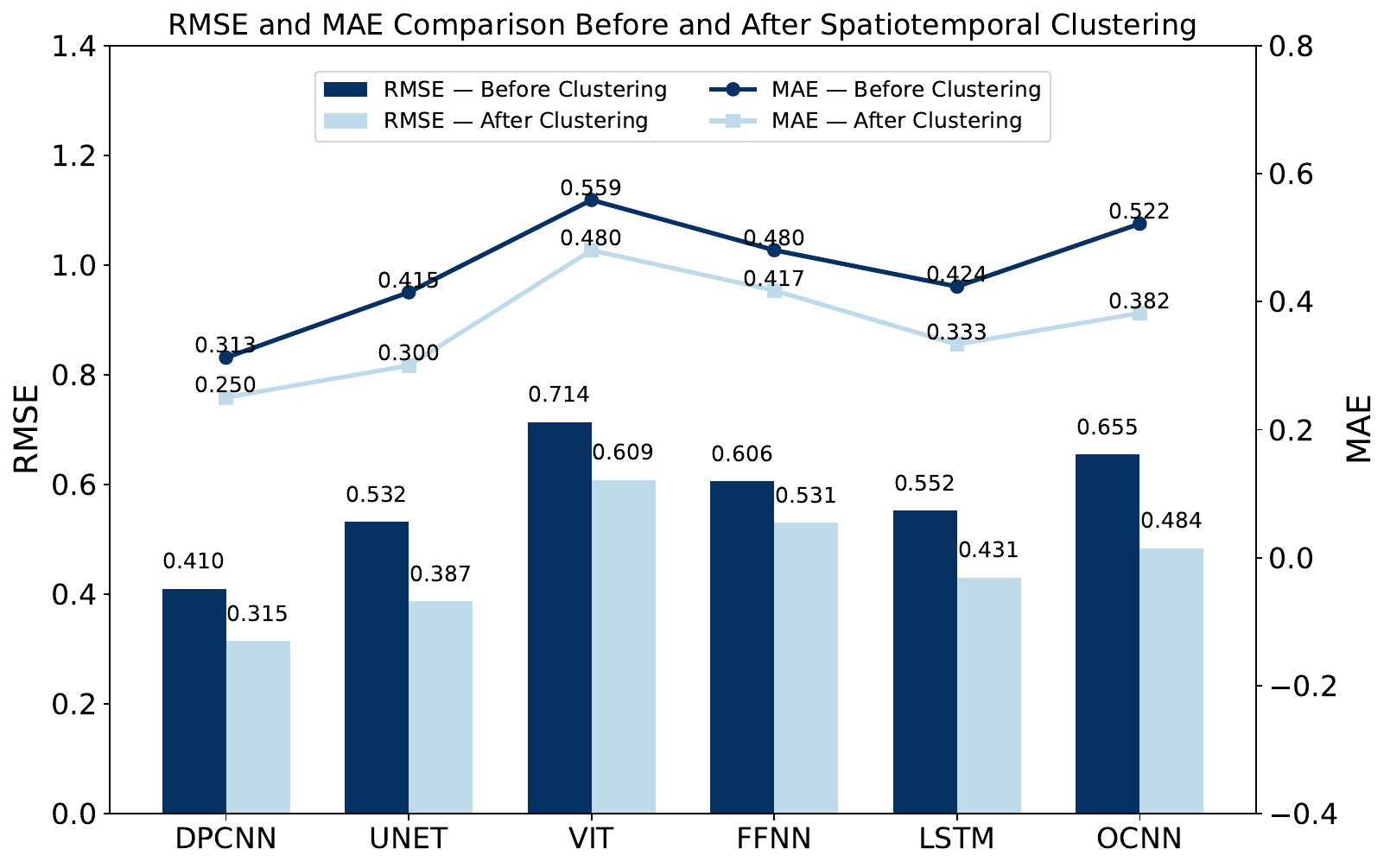}
        \caption{The South China Sea}
        \label{fig:rmse_mae_scs}
    \end{subfigure}\hfill
    \begin{subfigure}[t]{0.48\textwidth}
        \centering
        \includegraphics[width=\textwidth]{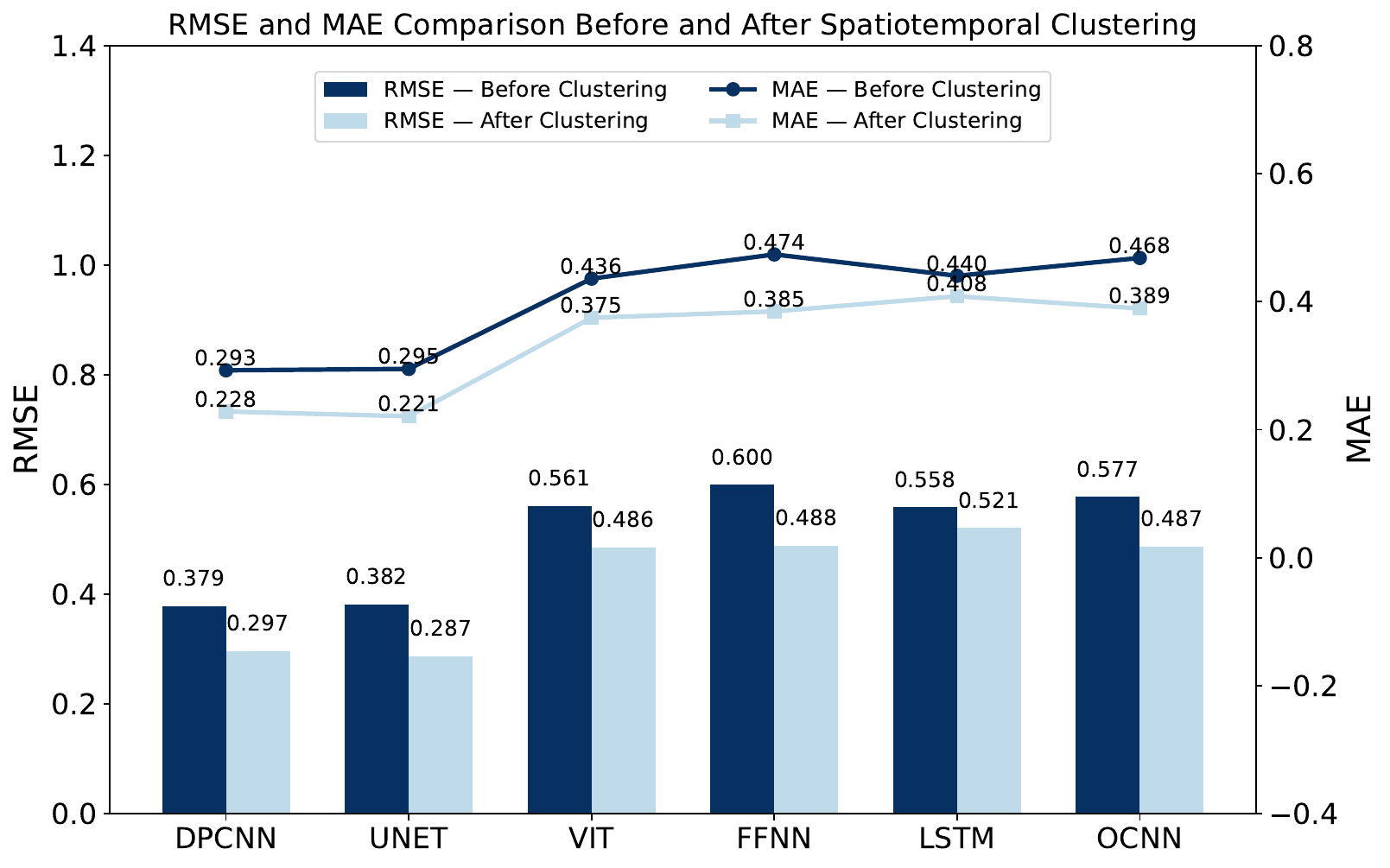}
        \caption{The Indian Ocean}
        \label{fig:rmse_mae_indian}
    \end{subfigure}

    \caption{
        Comparison of reconstruction errors before and after applying the proposed spatiotemporal clustering framework in two study areas.
        The bar plots denote RMSE values, while the line plots denote the corresponding MAE values.
    }
    \label{fig:rmse_comparison_dual}
\end{figure*}

\subsection{Comparison Results}
We conducted comparative experiments using six representative neural network backbones (DP-CNN, Attention U-Net, ViT, FFNN, LSTM, and OCNN) across two study areas. As shown in Figure~\ref{fig:rmse_comparison_dual}, both RMSE and MAE consistently decrease after applying spatiotemporal clustering in both areas, demonstrating the effectiveness of the proposed approach in mitigating spatiotemporal heterogeneity and enhancing reconstruction accuracy.

\subsection{Vertical and Temporal Clustering}

\begin{figure}[!htbp]
  \centering
  \captionsetup[sub]{font=small}
  \begin{subfigure}[t]{0.49\columnwidth}
    \centering
    \includegraphics[height=4cm,keepaspectratio]{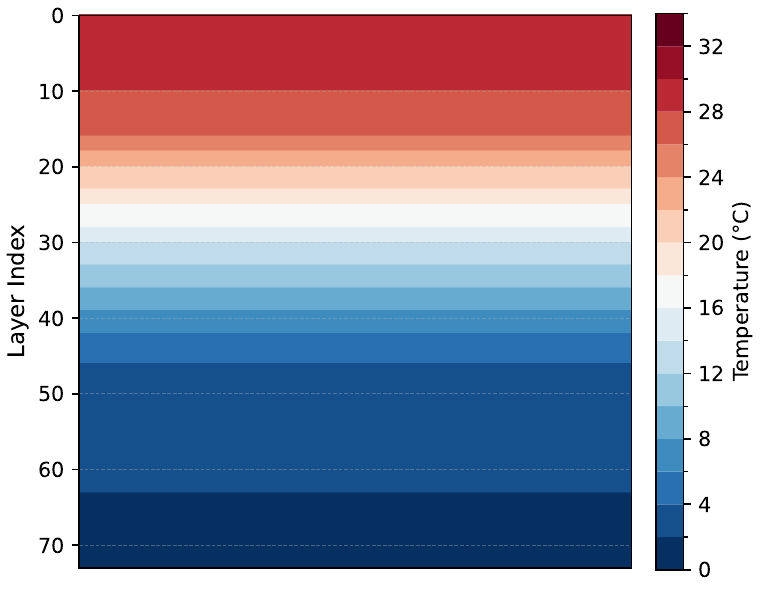}
    \caption{Temperature (layer index).}
    \label{fig:depth_A}
  \end{subfigure}\hfill
  \begin{subfigure}[t]{0.49\columnwidth}
    \centering
    \includegraphics[height=4cm,keepaspectratio]{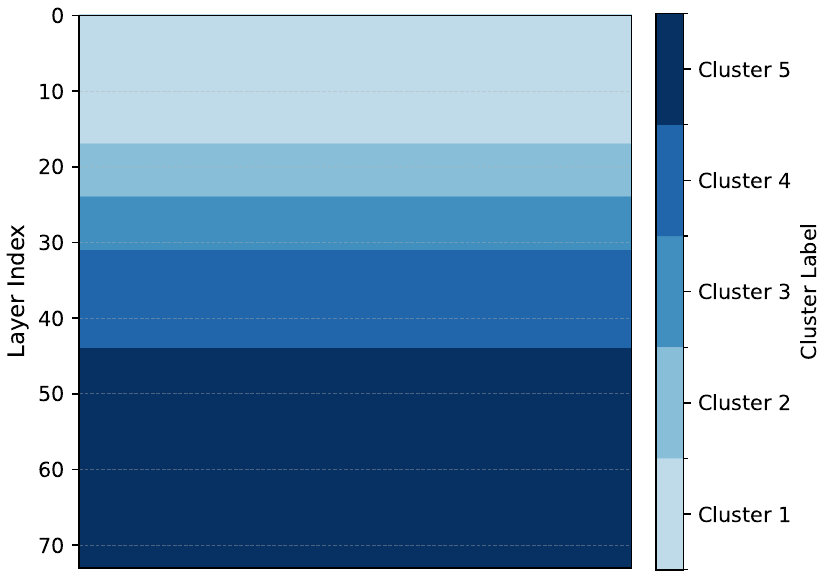}
    \caption{Clusters (layer index).}
    \label{fig:depth_B}
  \end{subfigure}
  \vspace{0.4em}
  \begin{subfigure}[t]{0.49\columnwidth}
    \centering
    \includegraphics[height=4cm,keepaspectratio]{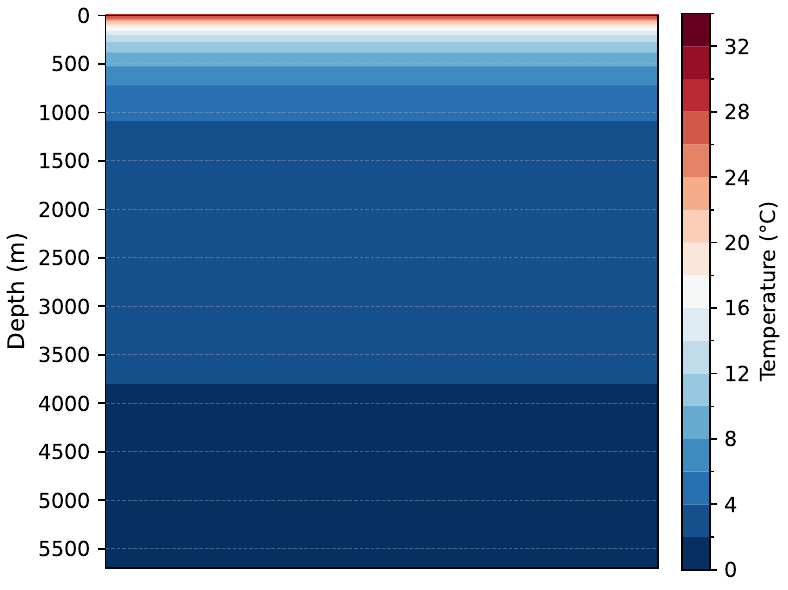}
    \caption{Temperature (real depth).}
    \label{fig:depth_C}
  \end{subfigure}\hfill
  \begin{subfigure}[t]{0.49\columnwidth}
    \centering
    \includegraphics[height=4cm,keepaspectratio]{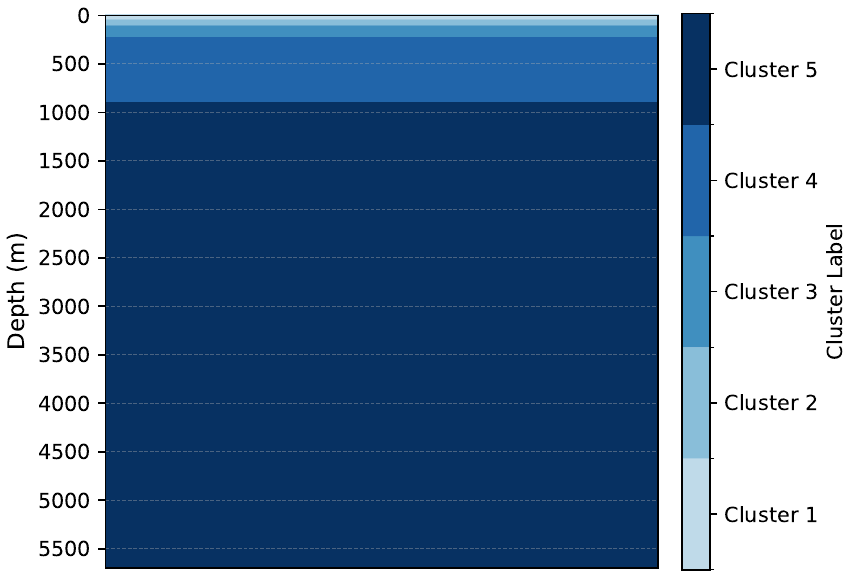}
    \caption{Clusters (real depth).}
    \label{fig:depth_D}
  \end{subfigure}
  \vspace{-0.3em}
  \caption{
    Vertical clustering results in the South China Sea.
    (a--b) Temperature section and cluster distribution along the layer index.
    (c--d) The corresponding results displayed along the real depth (m).
  }
  \label{fig:depth_clustering}
\end{figure}

As shown in the Figure~\ref{fig:depth_clustering}, the vertical clustering partitions the entire water column into several depth intervals, within which the temperature profiles exhibit strong statistical similarity and thermodynamic coherence. The near-surface interval is largely influenced by atmospheric forcing and seasonal mixing, the intermediate interval reflects changes in the thermocline and subsurface gradients, while the deeper layers maintain relatively stable thermal structures. This stratification not only highlights the physical differences across depth regimes but also reduces cross-layer heterogeneity, thereby facilitating more effective model learning.

\begin{figure}[!htbp]
  \centering
  \includegraphics[width=\columnwidth]{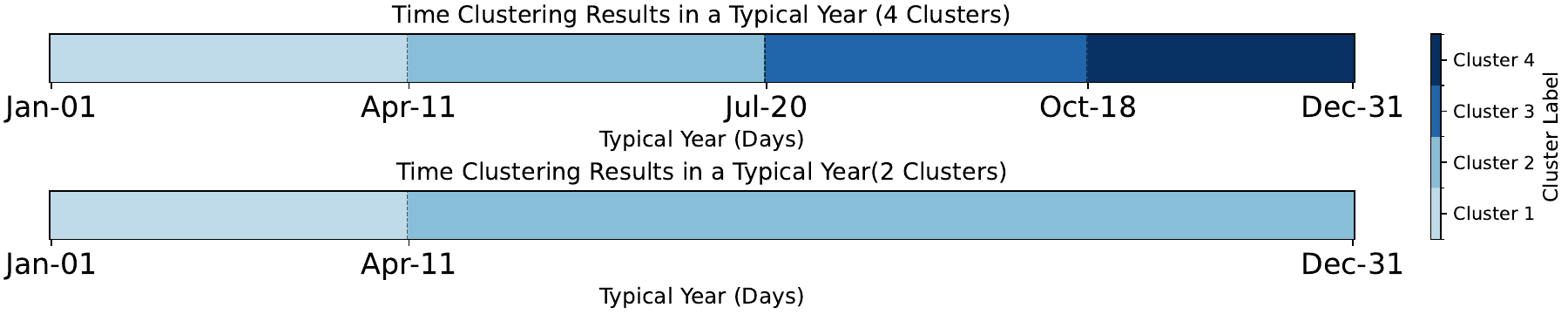}
  \caption{
    Temporal clustering results dividing the typical annual cycle into sub-periods corresponding to seasonal and climatic transitions.
    The 4-cluster scheme corresponds to the shallow-layer region (depth cluster~1), while the 2-cluster scheme corresponds to the deep-layer region (depth cluster~5).
  }
  \label{fig:time_clustering}
\end{figure}

 Figure~\ref{fig:time_clustering} illustrates the temporal clustering results for a typical year, where the annual cycle is divided into two and four clusters, respectively.
For regions with greater depths, the temporal variations are relatively smoother, and the annual cycle is divided into two clusters representing the broad warm and cool phases.
In contrast, some shallower regions exhibit more pronounced seasonal dynamics, leading to a finer four-cluster division that distinguishes the early-year warming, mid-year stable summer, autumnal transition, and late-year cooling phases.
This depth-dependent clustering effectively captures both the general annual cycle in deeper layers and the more complex seasonal transitions in surface waters.

\subsection{Ablation Study}
To evaluate the impact of clustering strategies on 3D ocean temperature reconstruction, ablation studies were conducted with six models: DP-CNN, Attention U-Net, ViT, FFNN, LSTM, and OCNN. As shown in Table~\ref{tab:clustering-rmse-mae}, all models yield the highest RMSE without clustering (e.g., 0.4097 for DP-CNN and 0.7140 for ViT). Vertical or temporal clustering alone provids moderate improvements, while the joint spatiotemporal clustering consistently achieves the best performance, reducing RMSE to 0.3151, 0.3874, 0.6087, 0.5306, 0.4308, and 0.4837 for DP-CNN, U-Net, ViT, FFNN, LSTM, and OCNN, respectively. These results confirm the effectiveness of joint clustering in mitigating spatiotemporal heterogeneity and enhancing generalization.

\begin{table*}[htbp]\rmfamily
    \centering
    \caption{Ablation results of clustering strategies for different models (RMSE / MAE)}
    \resizebox{\textwidth}{!}{
    \begin{tabular}{l*{6}{cc}}
        \toprule
        \textbf{Clustering Strategy} & 
        \multicolumn{2}{c}{\textbf{DP-CNN}} & 
        \multicolumn{2}{c}{\textbf{U-Net}} & 
        \multicolumn{2}{c}{\textbf{ViT}} &
        \multicolumn{2}{c}{\textbf{FFNN}} &
        \multicolumn{2}{c}{\textbf{LSTM}} &
        \multicolumn{2}{c}{\textbf{OCNN}} \\
        \cmidrule(lr){2-3}
        \cmidrule(lr){4-5}
        \cmidrule(lr){6-7}
        \cmidrule(lr){8-9}
        \cmidrule(lr){10-11}
        \cmidrule(lr){12-13}
        & RMSE & MAE & RMSE & MAE & RMSE & MAE & RMSE & MAE & RMSE & MAE & RMSE & MAE \\
        \midrule
        No Clustering         & 0.4097 & 0.3126 & 0.5320 & 0.4146 & 0.7140 & 0.5589 & 0.6059 & 0.4805 & 0.5521 & 0.4236 & 0.6546 & 0.5218 \\
        Vertical Clustering Only     & 0.3637 & 0.2893 & 0.4711 & 0.3680 & 0.6946 & 0.5516 & 0.5825 & 0.4632 & 0.4491 & 0.3485 & 0.5125 & 0.4062 \\
        Temporal Clustering Only     & 0.3673 & 0.2931 & 0.4832 & 0.3724 & 0.6323 & 0.4911 & 0.5563 & 0.4285 & 0.4438 & 0.3433 & 0.6247 & 0.5015 \\
        Joint Spatiotemporal Clustering   & \textbf{0.3151} & \textbf{0.2504} & \textbf{0.3874} & \textbf{0.3003} & \textbf{0.6087} & \textbf{0.4804} &
        \textbf{0.5306} & \textbf{0.4175} & \textbf{0.4308} & \textbf{0.3333} & \textbf{0.4837} & \textbf{0.3822} \\
        \midrule
        \textbf{Improvement (\%)} & \textbf{23.1\%} & \textbf{19.9\%} & \textbf{27.2\%} & \textbf{27.6\%} & \textbf{14.7\%} & \textbf{14.0\%} &
        \textbf{12.4\%} & \textbf{13.1\%} & \textbf{22.0\%} & \textbf{21.3\%} & \textbf{26.1\%} & \textbf{26.7\%} \\
        \bottomrule
    \end{tabular}
    }
    \label{tab:clustering-rmse-mae}
\end{table*}

\begin{figure*}[htbp]
  \centering
  \begin{subfigure}[b]{0.32\textwidth}
    \centering
    \includegraphics[width=\textwidth]{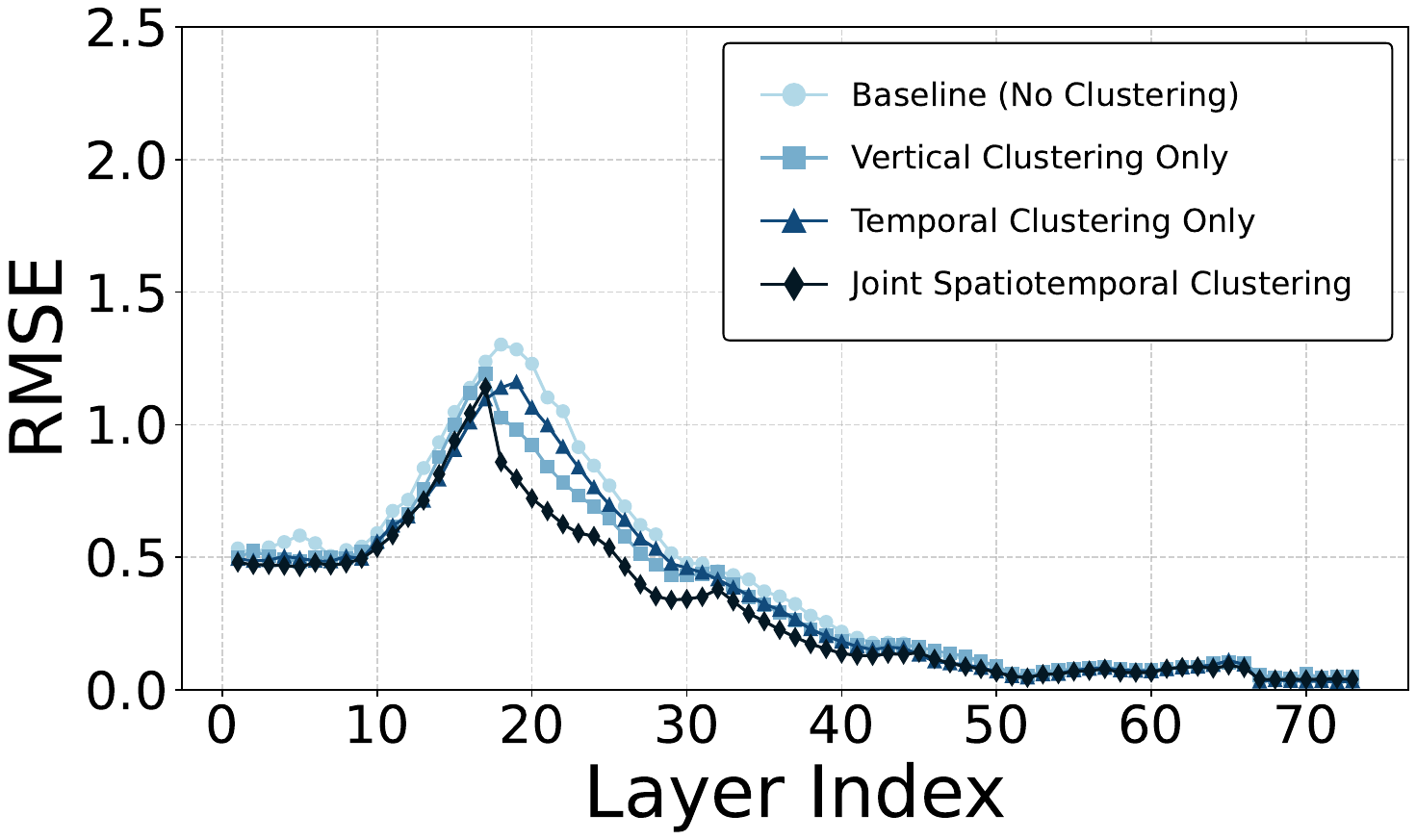}
    \caption{DP-CNN}
    \label{fig:rmse_dp_cnn}
  \end{subfigure}
  \hfill
  \begin{subfigure}[b]{0.32\textwidth}
    \centering
    \includegraphics[width=\textwidth]{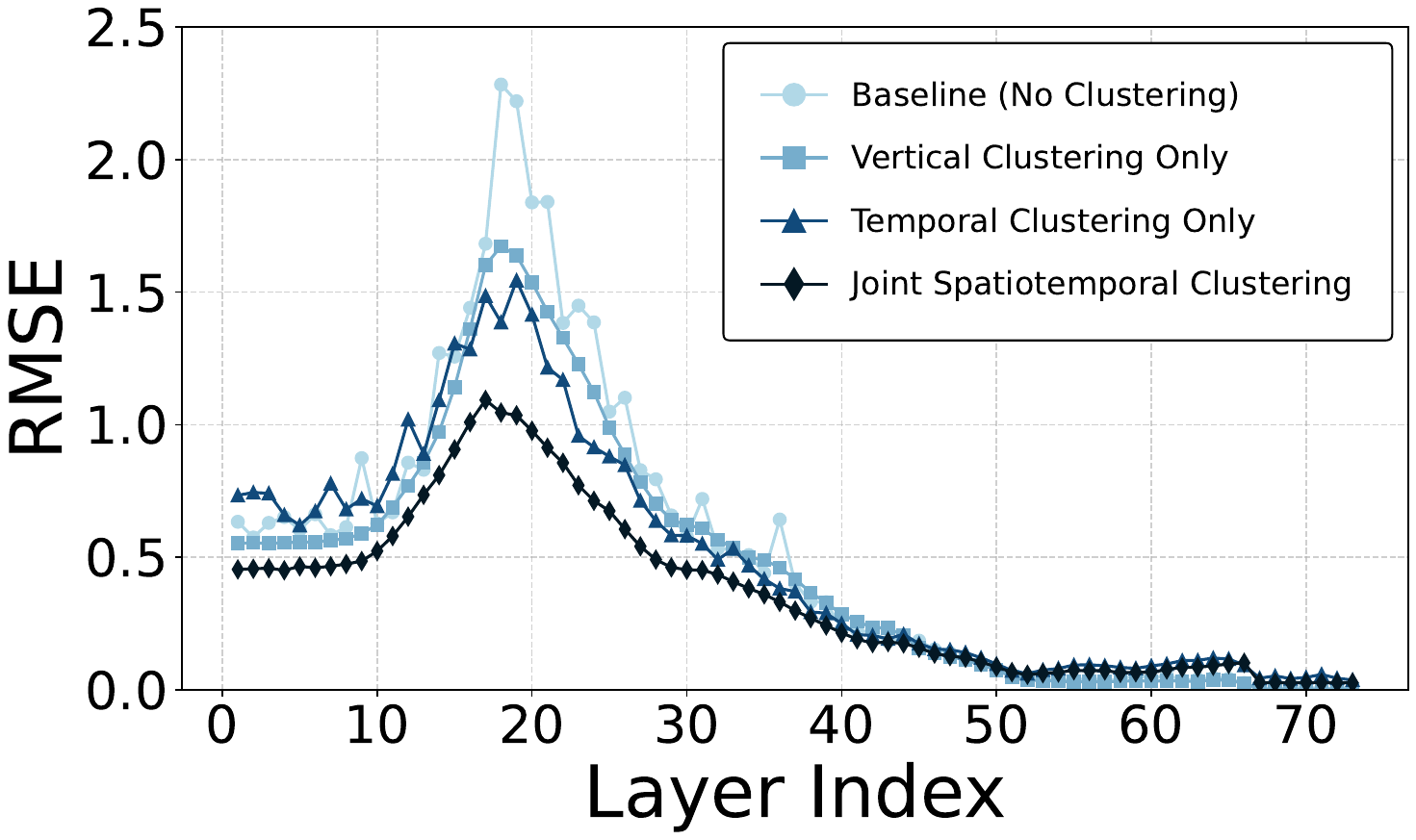}
    \caption{Attention U-Net}
    \label{fig:rmse_unet}
  \end{subfigure}
  \hfill
  \begin{subfigure}[b]{0.32\textwidth}
    \centering
    \includegraphics[width=\textwidth]{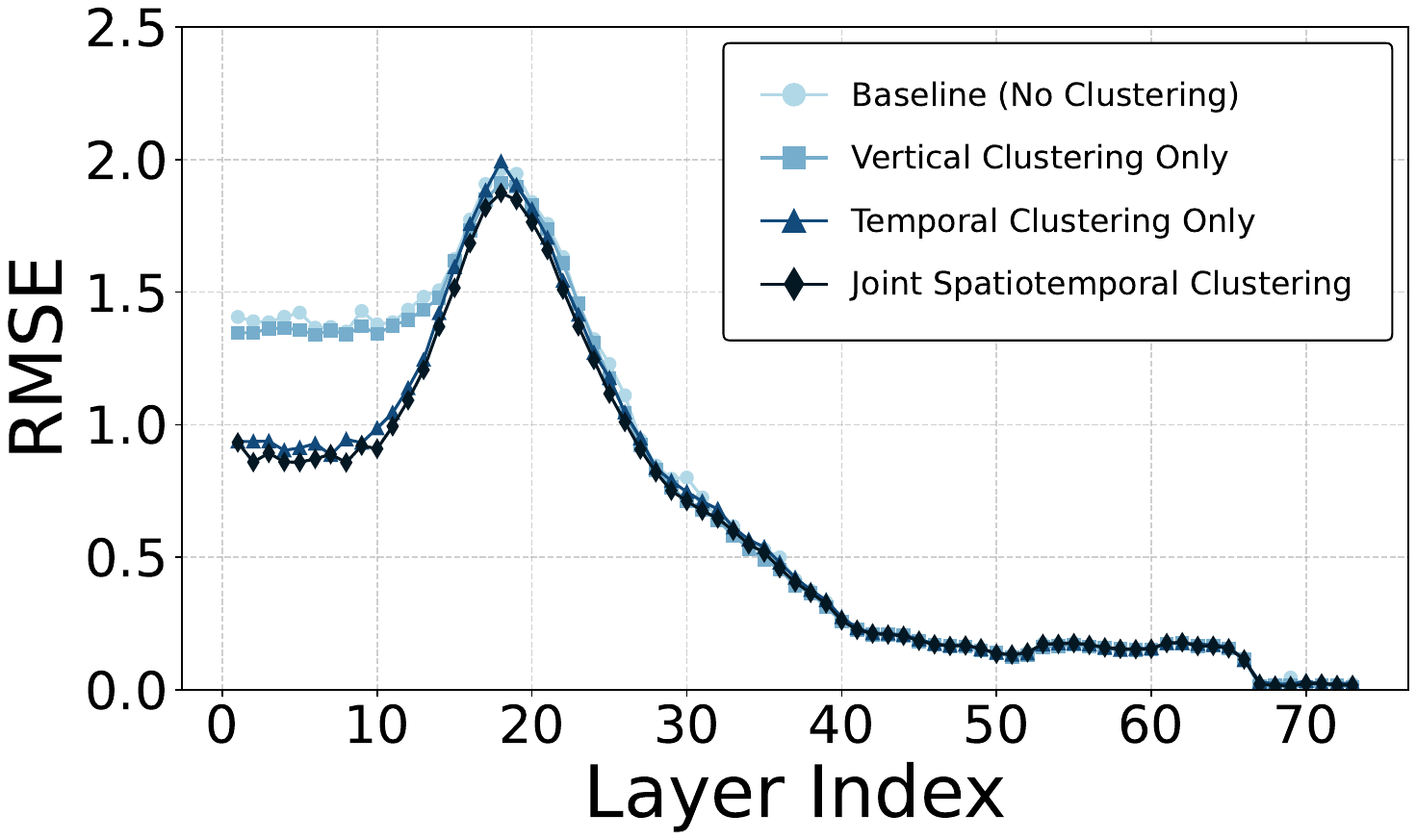}
    \caption{ViT}
    \label{fig:rmse_vit}
  \end{subfigure}

  \vskip\baselineskip  

  \begin{subfigure}[b]{0.32\textwidth}
    \centering
    \includegraphics[width=\textwidth]{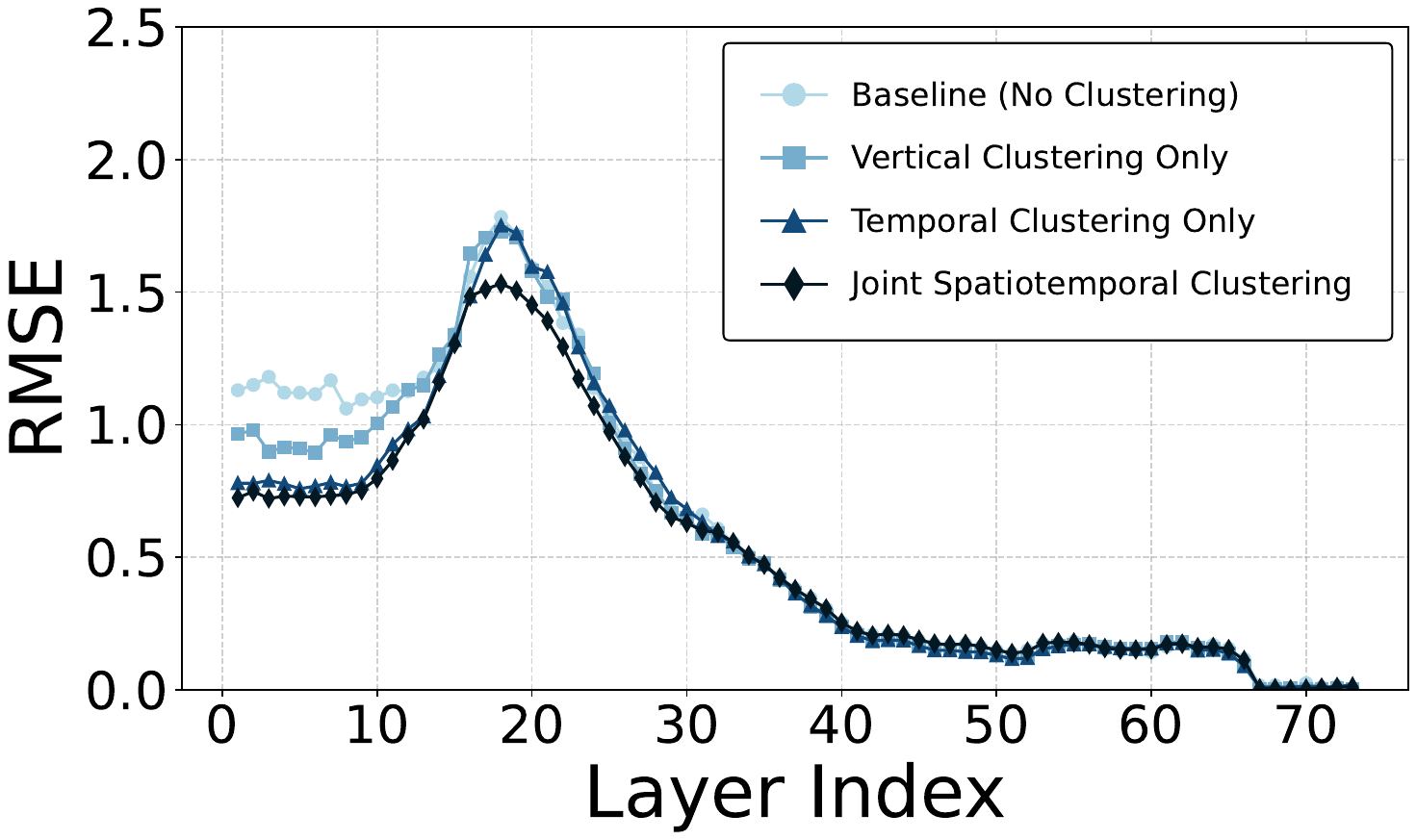}
    \caption{FFNN}
    \label{fig:rmse_ffnn}
  \end{subfigure}
  \hfill
  \begin{subfigure}[b]{0.32\textwidth}
    \centering
    \includegraphics[width=\textwidth]{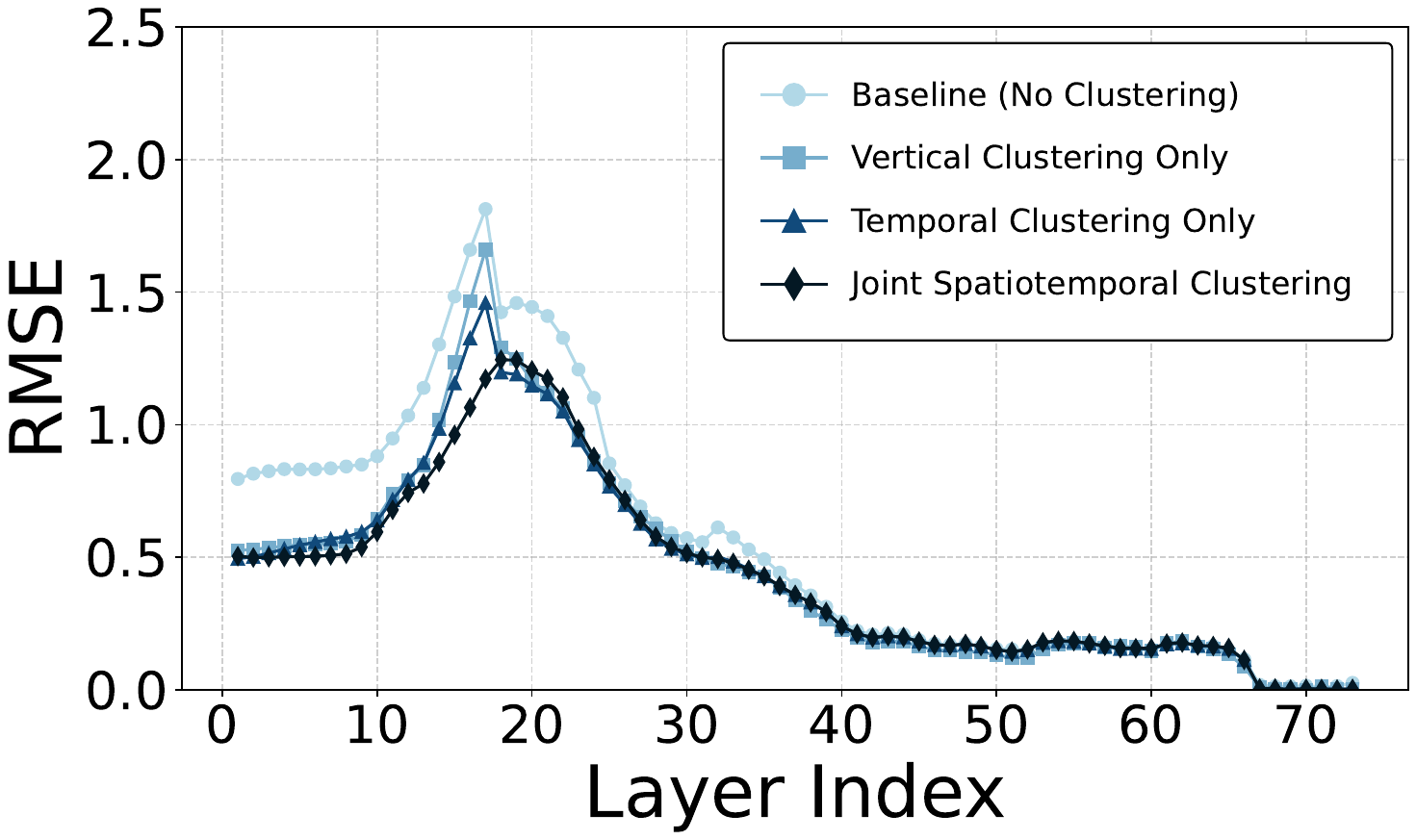}
    \caption{LSTM}
    \label{fig:rmse_lstm}
  \end{subfigure}
  \hfill
  \begin{subfigure}[b]{0.32\textwidth}
    \centering
    \includegraphics[width=\textwidth]{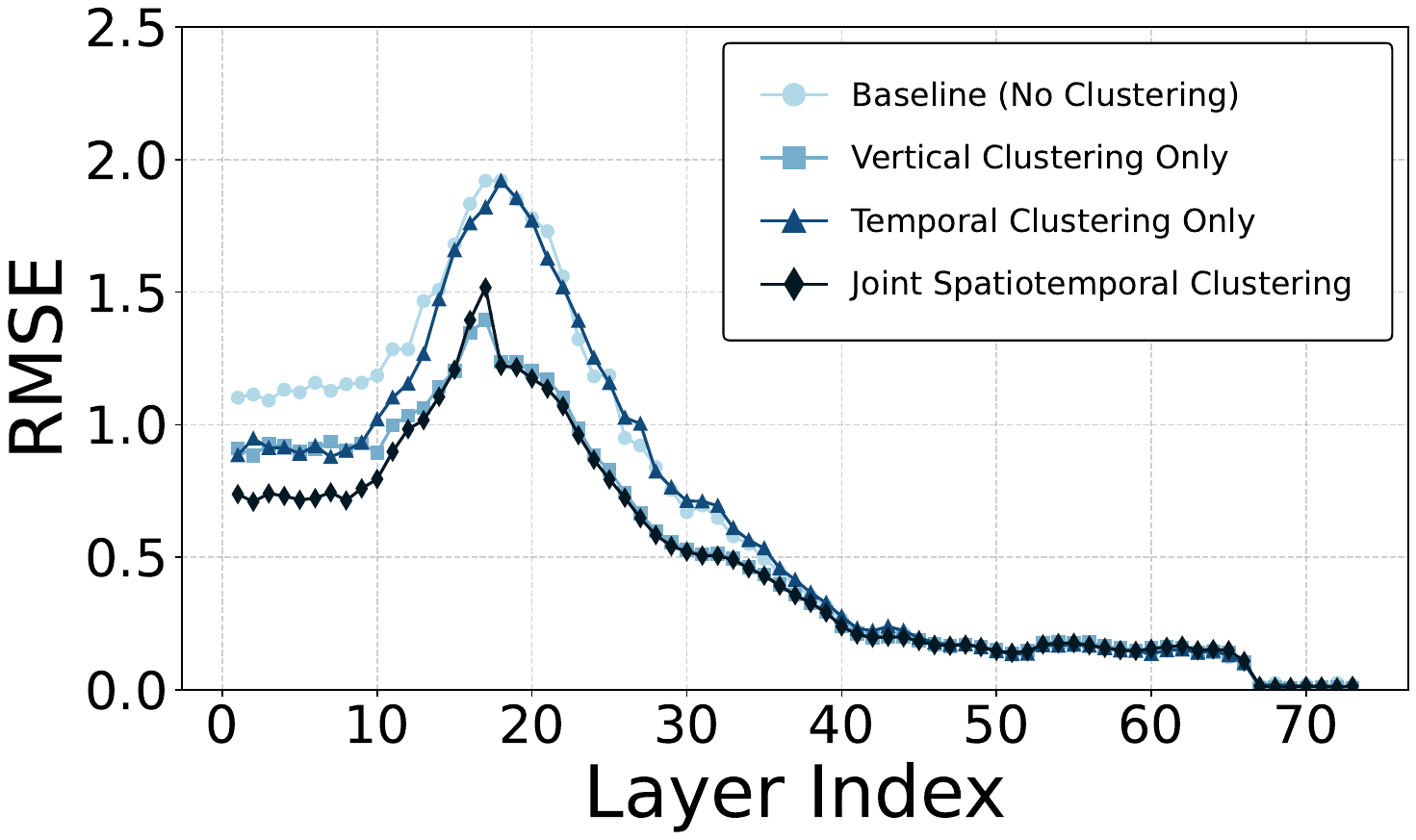}
    \caption{OCNN}
    \label{fig:rmse_ocnn}
  \end{subfigure}

  \caption{Layer-wise RMSE under different clustering strategies in the South China Sea. 
  }
  \label{fig:rmse_clustering_all}
\end{figure*}

Figure~\ref{fig:rmse_clustering_all} shows the layer-wise RMSE profiles under different clustering strategies.
All models exhibit a distinct error peak in the upper-middle depth layers (approximately layers 15-25) under the baseline (no clustering), indicating that the vertical thermal structure in this region is particularly complex and difficult to infer from surface observations.

Under the joint spatiotemporal clustering strategy, all models achieve their lowest RMSE across most depth layers, with smoother profiles. This confirms the effectiveness of combining vertical and temporal segmentation in modeling localized thermal structures. These results demonstrate that joint clustering mitigates cross-region interference and enhances the overall 3D temperature reconstruction performance.
\begin{figure*}[!t]  
  \centering
  \includegraphics[width=1\textwidth]{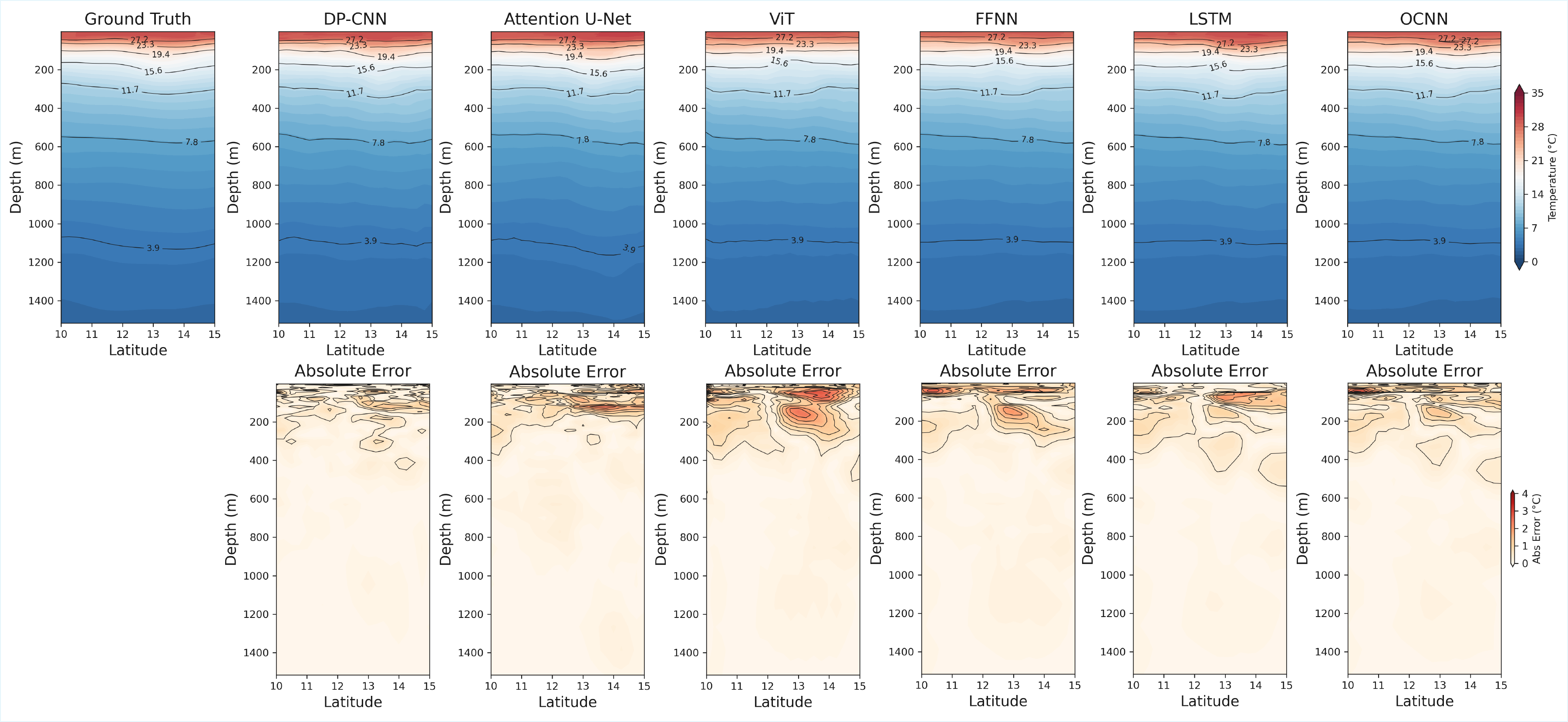} 
  \caption{Vertical profiles of reconstructed temperature fields within the upper 1500 meters and the corresponding absolute errors along the 112.5° latitude section on 24 August 2015.}
  \label{fig:profile_2015}
\end{figure*}

\begin{figure*}
  \centering
  \includegraphics[width=1\textwidth]{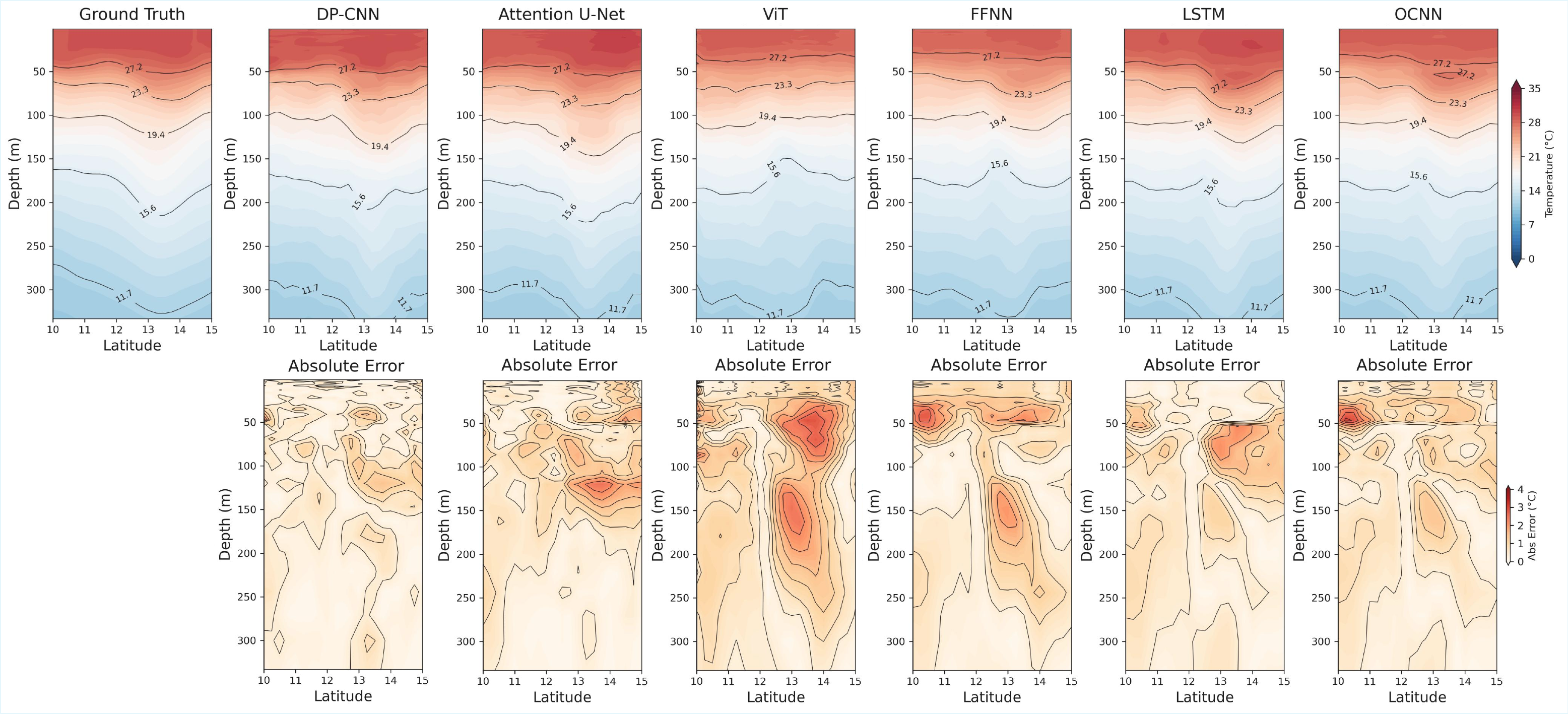} 
  \caption{Detailed comparison of reconstructed temperature fields within the upper 350 meters along the 112.5° latitude section on 24 August 2015, corresponding to the region shown in Figure~\ref{fig:profile_2015}. }
  \label{fig:profile_300}
\end{figure*}

\section{Visualization and Discussion}\label{sec:Visualization and Discussion}
As shown in Figure~\ref{fig:profile_2015}, all six deep learning models are able to reproduce the large-scale subsurface temperature structure along the 112.5° cross-section on 24 August 2015, but their accuracies differ considerably. The DP-CNN result shows the closest agreement with the ground truth, successfully recovering the layered thermocline structure with only minor deviations. The Attention U-Net also captures the stratified patterns, though its errors are slightly larger, particularly in the upper 200-400 m where the vertical gradients are strong. The ViT and FFNN models exhibit more pronounced deviations, with noticeable misrepresentation of sharp thermal transitions around the main thermocline. The LSTM yields moderate skill but still shows localized mismatches in regions of rapid temperature change. The OCNN achieves a performance comparable to DP-CNN, demonstrating robustness in representing both the near-surface variability and the deeper stable layers.

\begin{figure*}
    \centering
    \includegraphics[width=1\textwidth]{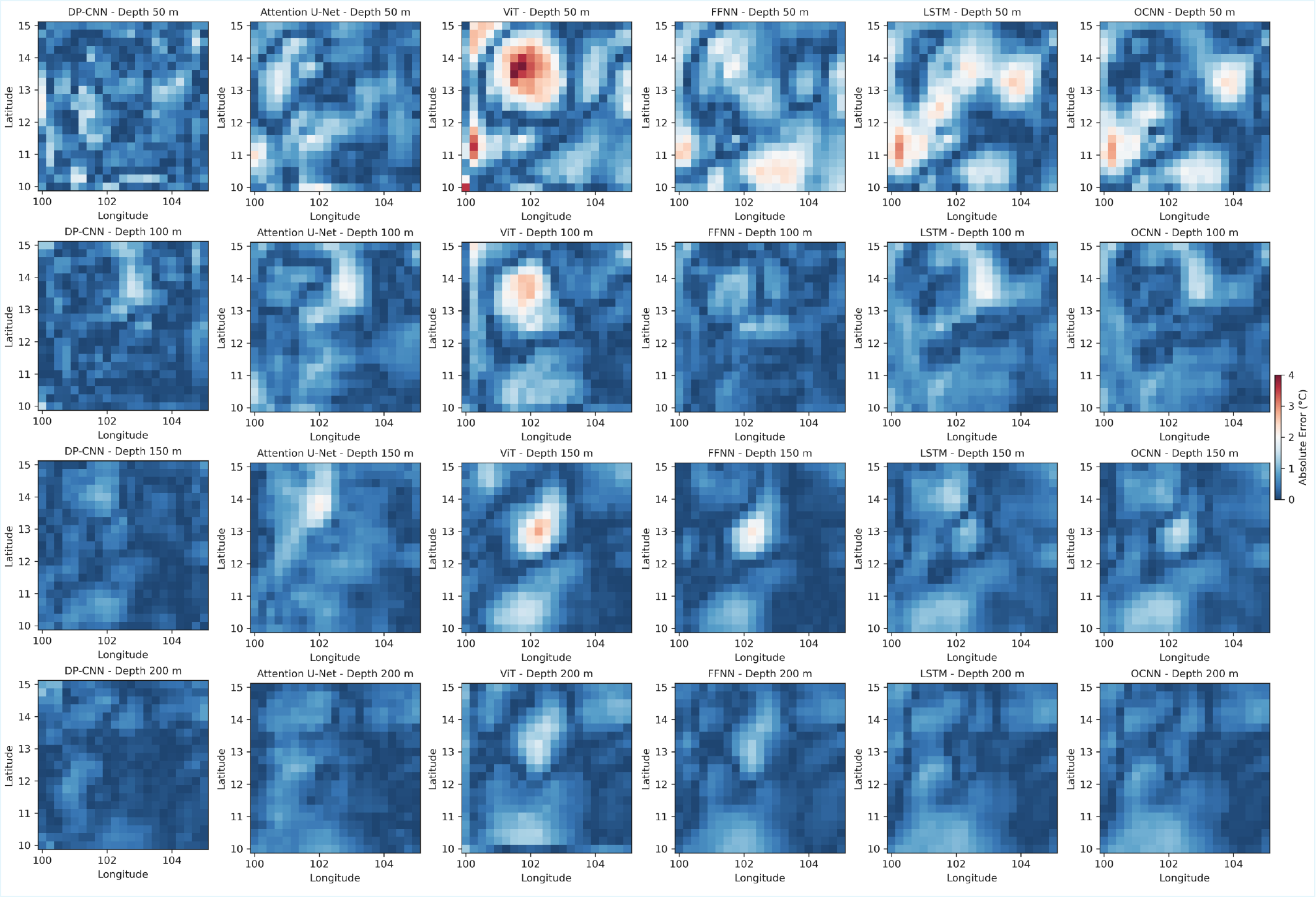} 
    \caption{Absolute error distributions at depths of 50 m, 100 m, 150 m, and 200 m.}
    \label{fig:heatmaps}
\end{figure*}

The absolute error maps highlight a common challenge across all models: errors concentrate near the main thermocline, where the vertical temperature gradient is steepest and small misalignments lead to large discrepancies. Shallow layers dominated by atmospheric forcing and mixing also pose difficulty, as rapid fluctuations are harder to capture. In contrast, the deeper layers remain relatively stable and are generally well reconstructed by all models.

Figure~\ref{fig:profile_300} presents a comparison of the reconstructed subsurface temperature fields in the upper 300 meters along the 112.5° latitude section on 24 August 2015, corresponding to the region shown in Figure~\ref{fig:profile_2015}. The top row shows the ground truth together with reconstructions from six models,
while the bottom row illustrates the corresponding absolute error maps. Overall, all models are able to reproduce the large-scale vertical stratification, but their accuracies vary. DP-CNN and OCNN provide the most faithful reconstructions, with relatively small errors near the surface and thermocline. Attention U-Net also achieves reasonable performance, though with slightly larger deviations in the 100-200 m range. In contrast, ViT and FFNN exhibit more pronounced errors in regions of sharp vertical gradients, while LSTM yields moderate accuracy but still shows localized discrepancies. The absolute error maps highlight a common challenge: reconstruction errors are mainly concentrated in the upper 100-200 m, where strong stratification and rapid variability increase the difficulty of accurate prediction.

Figure~\ref{fig:heatmaps} presents heatmaps of the absolute error distributions for six models 
at depths of 50 m, 100 m, 150 m, and 200 m. Across all methods, errors generally decrease with increasing depth, as deeper layers exhibit more stable thermal structures. At 50 m and 100 m, the error patterns are more dispersed due to strong stratification and surface forcing. ViT shows localized high-error regions in the shallow layers, while FFNN and LSTM also display noticeable deviations in certain areas. In contrast, DP-CNN and OCNN achieve smaller and more evenly distributed errors across all depths, and Attention U-Net demonstrates relatively uniform and moderate errors in the upper layers. By 150-200 m, all models converge toward lower and spatially smoother error patterns. These findings suggest that convolution-based architectures (DP-CNN and OCNN) deliver more robust reconstructions across depths.

\section{Conclusion}\label{sec:Conclusion}
In this work, we proposed an adaptive spatiotemporal clustering framework for 3D ocean subsurface temperature reconstruction. By integrating vertical-dependency clustering and temporal-dynamics clustering, the framework adaptively partitions the 3D ocean field into coherent sub-blocks, thereby mitigating cross-layer heterogeneity,
enabling fine-grained representation of subsurface thermal structures, and facilitating more effective model training. 
Experimental results demonstrate that the proposed framework is general and extensible, supporting various backbones including DP-CNN, Attention U-Net, ViT, FFNN, LSTM, and OCNN models. Across various clustering strategies and architectures, the framework consistently delivers robust improvements. Overall, this study confirms the effectiveness of spatiotemporal clustering in 3D ocean reconstruction and provides a promising pathway for intelligent ocean sensing under sparse observational conditions. Future work will aim to develop a unified framework that jointly optimizes clustering and reconstruction, rather than treating them as separate stages, thereby further improving the robustness and accuracy of 3D ocean subsurface temperature reconstruction.

\acknowledgments
This work was supported in part by National Natural Science Foundation of China (No. 62202336).

%
%


%
%
%
%
%

\bibliography{references}

\end{document}